\documentclass[12pt]{article}
\usepackage[top=30truemm,bottom=25truemm,left=25truemm,right=25truemm]{geometry}
\usepackage[dvipdfmx]{graphicx}
\usepackage{comment}
\usepackage{mathrsfs}
\usepackage{wrapfig}
\usepackage{bm}
\usepackage{cite}
\usepackage{braket}
\def\u#1{\mathrm {#1}} 

\begin{document}
\begin{center}
{\LARGE Effects of Pnictogen Atmosphere Annealing on Fe$_{1+y}$Te$_{0.6}$Se$_{0.4}$}

\vspace{5mm}
Tatsuhiro Yamada, Yue Sun, Sunseng Pyon, and Tsuyoshi Tamegai

\vspace{5mm}
{\it Department of Applied Physics, The University of Tokyo, 7-3-1 Hongo, Bunkyo-ku, Tokyo 113-8656, Japan}

\vspace{5mm}
abstract
\end{center}

It has been clarified that bulk superconductivity in Fe$_{1+y}$Te$_{0.6}$Se$_{0.4}$ can be induced by annealing in an appropriate atmosphere to remove the harmful effects of excess iron. 
In order to clarify the details of the annealing process, we studied the changes in the physical properties and reaction products of Fe$_{1+y}$Te$_{0.6}$Se$_{0.4}$ annealed in pnictogen (P, As, Sb) atmospheres. 
Crystals annealed in a pnictogen atmosphere show bulk superconductivity and the values of $T_\u{c}$ and $J_\u{c}$ are about $14~$K and 2-4 $\times 10{^5}~$A/cm$^2$ ($2~$K, self-field), respectively. 
It is also found that the reaction rate increases with the increase in the saturated vapor pressure of the pnictogen.
Unexpectedly, the reaction products of Fe$_{1+y}$Te$_{0.6}$Se$_{0.4}$ after annealing in a P atmosphere mainly consist of FeTe$_2$. 
In addition, the amount of P required to obtain the optimal $T_\u{c}$ is much smaller than the amount of excess iron, which is similar to the case of oxygen annealing.
P, oxygen, and to some extent As could serve as catalysts to form FeTe$_2$ to remove excess iron.

\clearpage
\section{Introduction}
The discovery of superconductivity in LaFeAs(O,F) has attracted much attention to superconductors containing two-dimensional iron planes and has led to the discovery of many iron-based superconductors (IBSs)\cite{JAmChemSoc.130.3296}.
The superconducting mechanism of IBSs has been proposed to involve two possible scenarios, the $s_\pm$\cite{PhysRevLett.101.057003,PhysRevLett.101.087004} and $s_{++}$\cite{PhysRevLett.104.157001,PhysRevB.81.054518} scenarios, where the pairing interaction is mediated by spin and orbital fluctuations, respectively.
The actual scenario is still under intense debate because the multiband structures of IBSs, which are characterized by hole bands around the $\Gamma$ point and electron bands around the M point\cite{PhysRevLett.104.097002,PhysRevB.78.134514,PhysRevB.81.014526}, make it difficult to conclude satisfactorily which is responsible for superconductivity.
Among the IBSs, Fe$_{1+y}$Te$_{1-x}$Se$_{x}$ (the 11 system) has the simplest crystal structure, composed of only superconducting planes\cite{ProcNatlAcadSci.105.14262}, making it an ideal system to study the mechanism of superconductivity in IBSs. 
In addition, the 11 system is less toxic than other IBSs, which is also preferable for applications.
However, it is known that as-grown crystals of Fe$_{1+y}$Te$_{1-x}$Se$_{x}$ do not show superconductivity\cite{PhysRevB.82.212504} due to the presence of excess iron that causes electron localization, thereby serving as pair breakers\cite{PhysRevB.80.174509,PhysRevB.79.014522}. 

Efforts to remove excess iron have been made by annealing the crystal in various atmospheres, such as vacuum\cite{PhysRevB.80.092502,JPhysSocJpn.79.084711}, oxygen\cite{SupercondSciTechonol.26.015015,SolidStateCommun.152.1135,SupercondSciTechnol.25.084011}, Te\cite{JPhysSocJpn.82.093705,JPhysSocJpn.82.023703}, Se, S\cite{JPhysSocJpn.82.115002}, As\cite{JPhysSocJpn.83.064704}, and I\cite{SupercondSciTechonol.26.015015,JAmChemSoc.132.10006}.
Through these annealing processes, excess iron existing in the second Fe site\cite{PhysRevLett.102.247001,JPhysSocJpn.82.023703} is removed\cite{SciRep.4.4585}, and the distribution of Te and Se becomes homogeneous\cite{PhysRevB.80.092502}, leading to bulk superconductivity, which has been confirmed by specific heat measurements\cite{JPhysSocJpn.81.054708,PhysRevB.83.134521}. 
Recently, it has been reported that FeTe$_m$ complexes are formed by annealing Fe$_{1.05}$Te$_{0.75}$Se$_{0.25}$ in Te vapor.
This degrades the superconducting properties around FeTe$_m$ because FeTe$_m$ is a magnetic impurity similar to the excess iron\cite{PhysRevB.91.060513}.
On the other hand, the optimal molar ratio of oxygen to the crystal to obtain the optimal transition temperature, $T_\u{c}$, is 1.5{\%}\cite{SciRep.4.4585}, while it is about 10-20{\%} in the case of Te annealing\cite{JPhysSocJpn.82.093705}.
The origin of this discrepancy is not yet understood clearly, because the details of these annealing processes including their dynamics are not fully understood. 
By comparing different annealing processes, we can optimize the annealing conditions to remove harmful excess iron between Te/Se planes without disturbing the crystal structure. 
This will help to obtain high-quality single crystals and may allow us to improve the critical current density, $J_\u{c}$, of a wire of the 11 system, which is reported to be only $\sim 750~$A/cm$^2$\cite{SupercondSciTechonol.24.075025}. 
For these reasons, it is important to unveil the details of the annealing process. 
Comparing the effects of annealing in various atmospheres can give us an insight into which atmosphere is the most suitable for improving the quality of single crystals.
In this paper, changes in the physical properties of Fe$_{1+y}$Te$_{0.6}$Se$_{0.4}$ after annealing in pnictogen atmospheres and its reactions are studied and compared with the case of oxygen annealing\cite{SciRep.4.4585}.

\section{Experimental Procedure}
Single crystals of Fe$_{1+y}$Te$_{0.6}$Se$_{0.4}$ were grown using the standard technique employing slow cooling\cite{RepProgPhys.74.124503}. 
As-grown crystals were cut into small pieces with rectangular shapes and annealed in various pnictogen atmospheres. 
The amounts of pnictogen and the crystal sealed in a quartz tube were controlled so that the molar ratio of the pnictogen to the crystal was in the range of 0.5-150{\%}. 
As-grown crystals and pnictogen shots were loaded into a quartz tube, which had been washed and baked carefully to remove impurities. 
The quartz tube was evacuated by an oil-diffusion pump. 
The annealing temperature was fixed to 400$~^\circ$C, and its time was varied between 15 min and 7 days, which was followed by quenching in water.
To evaluate the value for $T_\u{c}$ and $J_\u{c}$ of the annealed crystal, magnetization measurements were performed using a commercial superconducting quantum interference device magnetometer (MPMS-XL5, Quantum Design). 
Reaction products on the sample surface were identified by X-ray diffraction (XRD) to investigate the difference among the annealing processes in different pnictogen atmospheres.
Distributions of elements on the surface layers after annealing were confirmed by a scanning electron microscope (SEM) equipped with an energy-dispersive X-ray spectroscopy (EDX).
We also prepared oxygen-annealed crystal to compare oxygen annealing with the pnictogen annealing. 
Details of the oxygen annealing are described in Ref. 15.

\section{Results and Discussion}
\begin{figure*}[t]
\begin{center}
\includegraphics[width=\hsize]{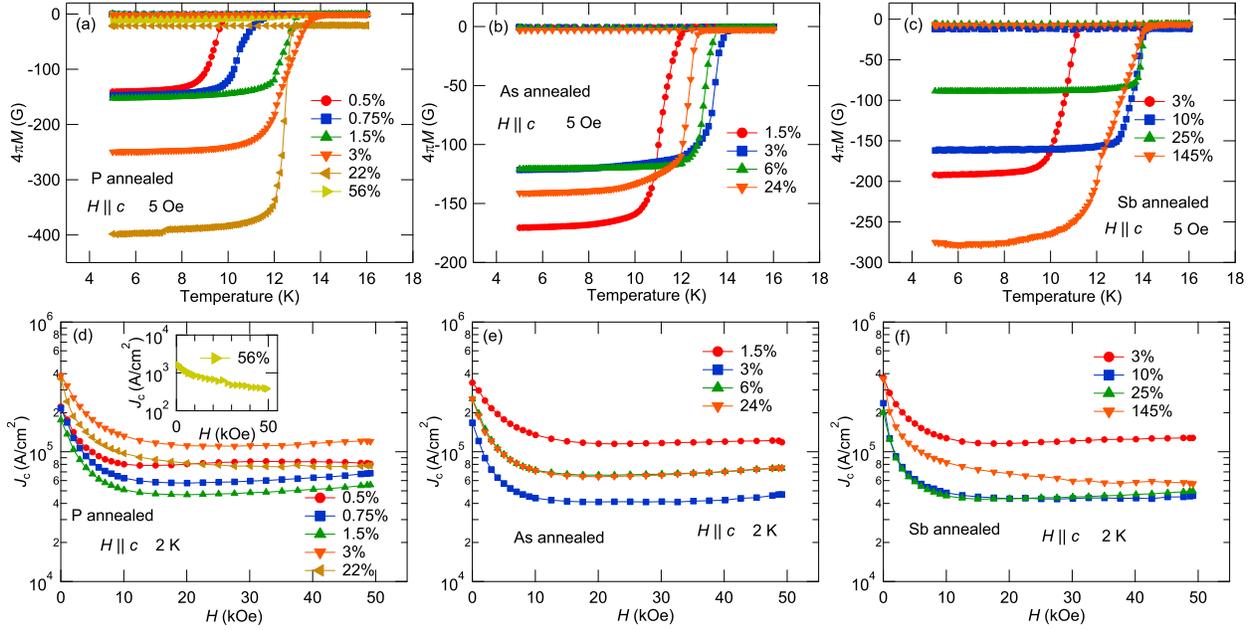}\\
\caption{(color online) Selected temperature dependences of zero-field-cooled (ZFC) and field-cooled (FC) magnetization at $5~$Oe for Fe$_{1+y}$Te$_{0.6}$Se$_{0.4}$ annealed in (a) P, (b) As, and (c) Sb atmospheres. Magnetic field dependence of $J_\u{c}$ for Fe$_{1+y}$Te$_{0.6}$Se$_{0.4}$ annealed in (d) P, (e) As, and (f) Sb atmospheres.}
\label{TcJc}
\end{center}
\end{figure*}
Figures \ref{TcJc}(a)-\ref{TcJc}(c) show selected temperature dependences of zero-field-cooled (ZFC) and field-cooled (FC) magnetization at $5~$Oe for Fe$_{1+y}$Te$_{0.6}$Se$_{0.4}$ crystals annealed for about 48 hours in P, As, and Sb atmospheres, respectively. 
Molar ratios written in the figures are the ratio of the element to the crystal.
We define $T_\u{c}$ by the points of intersection between the FC curve and the line extrapolated from the ZFC curve.
The different values of magnetization at low temperatures are due to the different demagnetization effect for each crystal.
Figures \ref{TcJc}(d)-\ref{TcJc}(f) show $J_\u{c}$ as a function of magnetic field for Fe$_{1+y}$Te$_{0.6}$Se$_{0.4}$ corresponding to Figs. \ref{TcJc}(a)-\ref{TcJc}(c), respectively.
The extended Bean model is used for the calculation of $J_\u{c}$ from  magnetic hysteresis loops\cite{RevModPhys.36.31}.
$T_\u{c}$ for the crystals annealed in each pnictogen atmosphere increases as the amount of pnictogen increases and reaches the maximum at the optimal ratio for each pnictogen. 
As the amount of pnictogen increases further, $T_\u{c}$ decreases gradually. 
Except for the case of P annealing, the optimal $T_\u{c}$ and $J_\u{c}$ are $\sim 14~$K and 2-4$\times10^{5}~$A/cm$^2$ ($2~$K, self-field), respectively.
These values are nearly equal to those of oxygen-annealed crystals\cite{SciRep.4.4585}. 
However, a closer inspection indicates that the maximum $T_\u{c}$ for P-annealed crystals is only $13.7~$K. 
The high $T_\u{c}$ and  $J_\u{c}$ comparable to the reported highest values means that the superconductivity in Fe$_{1+y}$Te$_{0.6}$Se$_{0.4}$ crystals annealed in P, As, and Sb atmospheres is bulk in nature.
As shown in Fig. \ref{TcJc}, the difference among the pnictogens can be seen clearly. 
In the case of P annealing, the optimal ratio of P to the sample is about $3~${\%}, which is similar to the case of oxygen annealing\cite{SciRep.4.4585,ApplPhysExpress.6.043101}. 
On the other hand, when the molar ratio of P increases to $56${\%}, the quality of the crystal deteriorates, as seen from the small absolute value of magnetization and small $J_\u{c}$. 
This is possibly due to the reaction between P and Fe$_{1+y}$Te$_{0.6}$Se$_{0.4}$ itself. 
In the case of As annealing, $T_\u{c}$ reaches an optimal value at around 3-6{\%}, which is similar to the case of P annealing.
It is worth noting that the reaction products peeled off from the surface of the crystal and they were clearly seen on the inner wall of the quartz tube when the molar ratio of As was more than $40${\%}.
These reaction products are compounds consisting of Fe and As\cite{JPhysSocJpn.83.064704}. 
In the case of Sb annealing, it is very surprising that $T_\u{c}$ and $J_\u{c}$ remain large even though the molar ratio of Sb is more than $100${\%}.
This may be due to the slow reaction between Sb and the crystals.
Actually, when the amount of Sb is more than $40${\%}, some of the starting Sb remains in the quartz tube after annealing.

\begin{figure}[htbp]
\begin{center}
\includegraphics[width=\hsize]{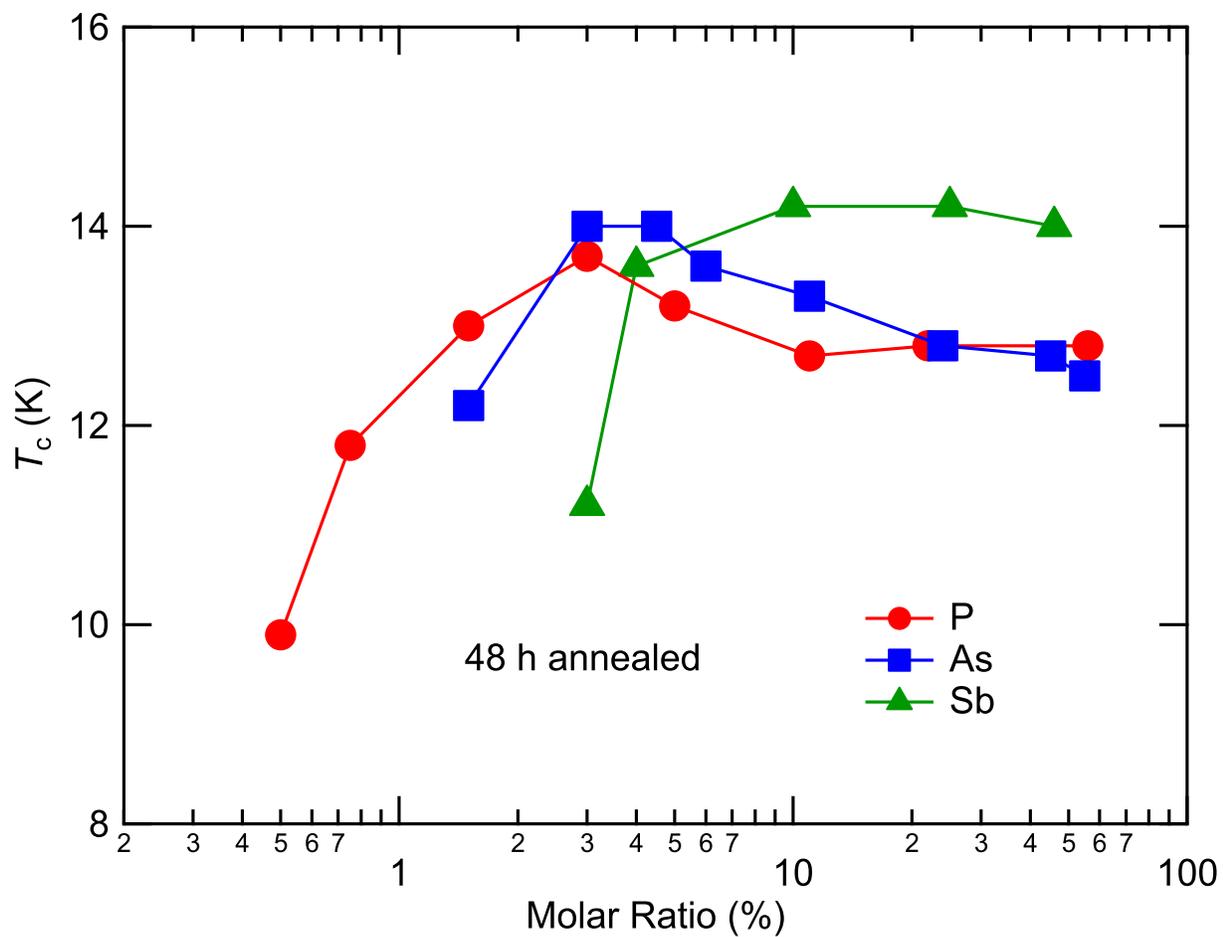}\\
\caption{(color online) $T_\u{c}$ as a function of the molar ratio for annealing in different pnictogens at 400$~^\circ$C for 48 h. Note that all the raw data are not shown in Fig. \ref{TcJc}.}
\label{Tccompare}
\end{center}
\end{figure}
Figure \ref{Tccompare} summarizes the molar ratio dependence of $T_\u{c}$ shown in Fig. 1. 
There is an ambiguity in the relationship between the amount of excess iron and the value of $J_\u{c}$ because impurities may serve as pinning centers for vortices and affect $J_\u{c}$. 
Namely, the amount of excess iron contributes to the value of $J_\u{c}$ more intricately, and it is more suitable to deal with the value of $T_\u{c}$. 
As stated earlier, the optimal molar ratio to obtain the highest $T_\u{c}$ is a few percent for P and As annealing and around 10{\%} for Sb annealing,  which are similar to those for oxygen and Te annealing, respectively.
We compare oxygen annealing with P annealing at the same molar ratio of 0.5{\%} to verify the similarity more quantitatively.
The values of $T_\u{c}$ and $J_\u{c}$ for the oxygen-annealed samples are $\sim 6~$K and less than $1 \times 10^5~$A/cm$^2$(Ref. 24), corresponding to $\sim 10~$K and $\sim 2 \times 10^5~$A/cm$^2$ in the P-annealed crystal (not shown here), respectively, indicating that P may also react with Fe$_{1+y}$Te$_{0.6}$Se$_{0.4}$ similarly to the case of oxygen.
In contrast to P and As annealing, $T_\u{c}$ remains high for the samples annealed in more than 20{\%} Sb.

\begin{figure}[htbp]
\begin{center}
\includegraphics[width=\hsize]{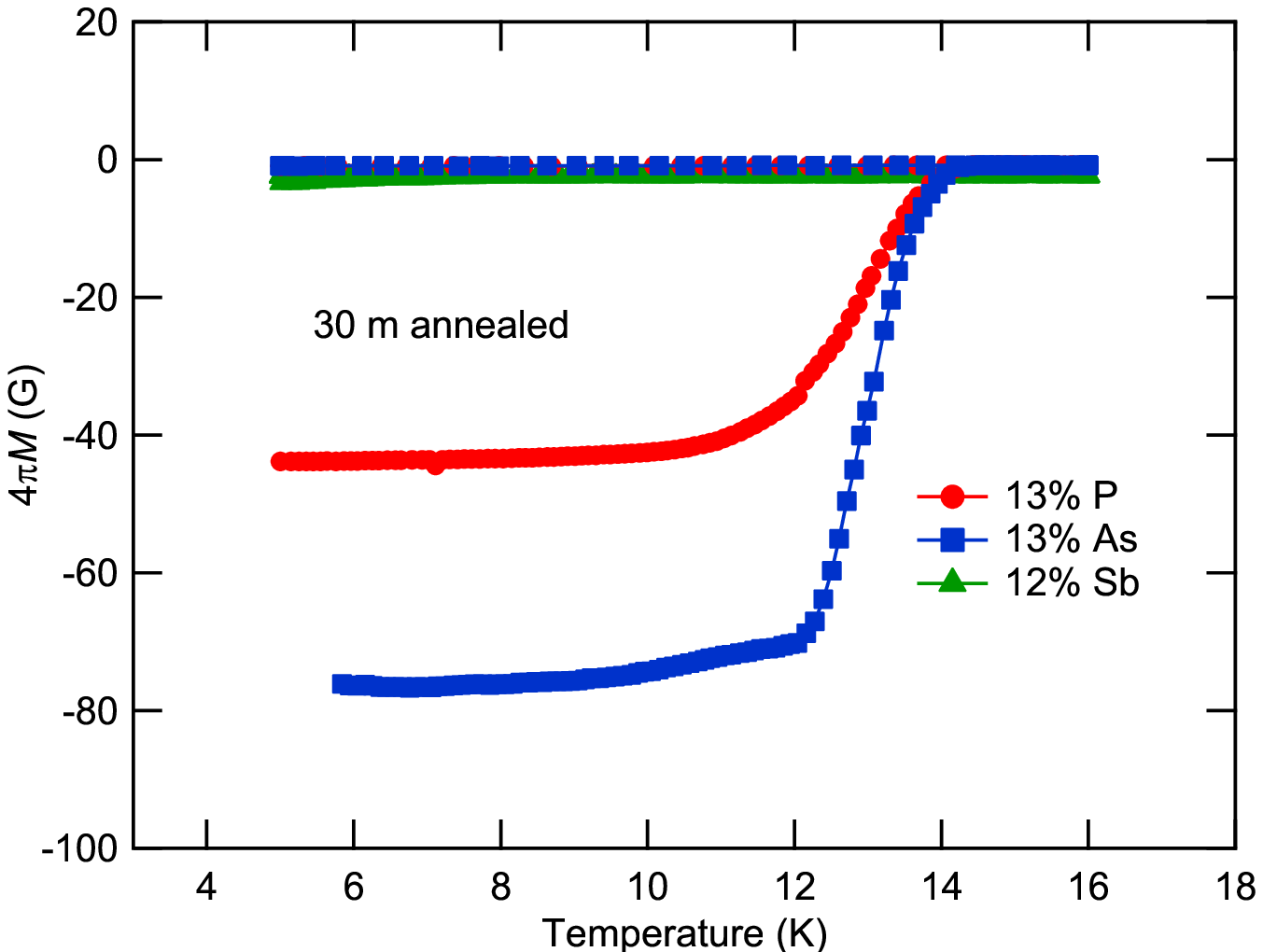}\\
\caption{(color online) Temperature dependence of ZFC and FC magnetization at $5~$Oe for Fe$_{1+y}$Te$_{0.6}$Se$_{0.4}$ annealed at 400$~^\circ$C for 30 min in 13{\%} P, 13{\%} As, and 12{\%} Sb atmospheres.}
\label{speed_compare}
\end{center}
\end{figure}
The differences observed in annealing in different pnictogen atmospheres may originate from the difference in the dynamics of the reaction in different atmospheres. 
To obtain more insight into the dynamics of the reaction, we annealed Fe$_{1+y}$Te$_{0.6}$Se$_{0.4}$ crystals in P, As, and Sb atmospheres for a much shorter time at 400$~^\circ$C. 
Figure \ref{speed_compare} shows the temperature dependence of ZFC and FC magnetization at $5~$Oe for Fe$_{1+y}$Te$_{0.6}$Se$_{0.4}$ annealed at 400$~^\circ$C for 30 min in 13{\%} P, 13{\%} As, and 12{\%} Sb atmospheres.
The crystal annealed in the Sb atmosphere shows no superconductivity, while significant diamagnetism develops below $\sim 14~$K in the crystals annealed in P and As atmospheres. 
It is natural to consider that the rate of the reaction is proportional to the saturated vapor pressure (SVP) at the annealing temperature. 
The SVP of Sb at 400$~^\circ$C is very low ($\sim 10^{-2}~$Pa) and not all the solid Sb can become a gas simultaneously\cite{Book}. 
We note that the SVPs of P and As are more than $100~$Pa at 400$~^\circ$C, and all the P and As can be vaporized since the vapor pressures are lower than the SVPs.
This fact can also explain why $T_\u{c}$ for the crystal annealed in an oxygen atmosphere for only $30$ min is also about $14~$K because the SVP of oxygen is much higher than that of the pnictogens\cite{SciRep.4.4585}. 
This is also the reason why the crystal annealed in a $145${\%} Sb atmosphere has a large $T_\u{c}$ and $J_\u{c}$. 
These results indicate that the optimal annealing time depends on the element because of the different SVPs. 
We anticipate a clear tendency that a longer annealing time is required to obtain the optimal $T_\u{c}$ in an atmosphere with a lower SVP.

\begin{figure}[htbp]
\begin{center}
\includegraphics[width=\hsize]{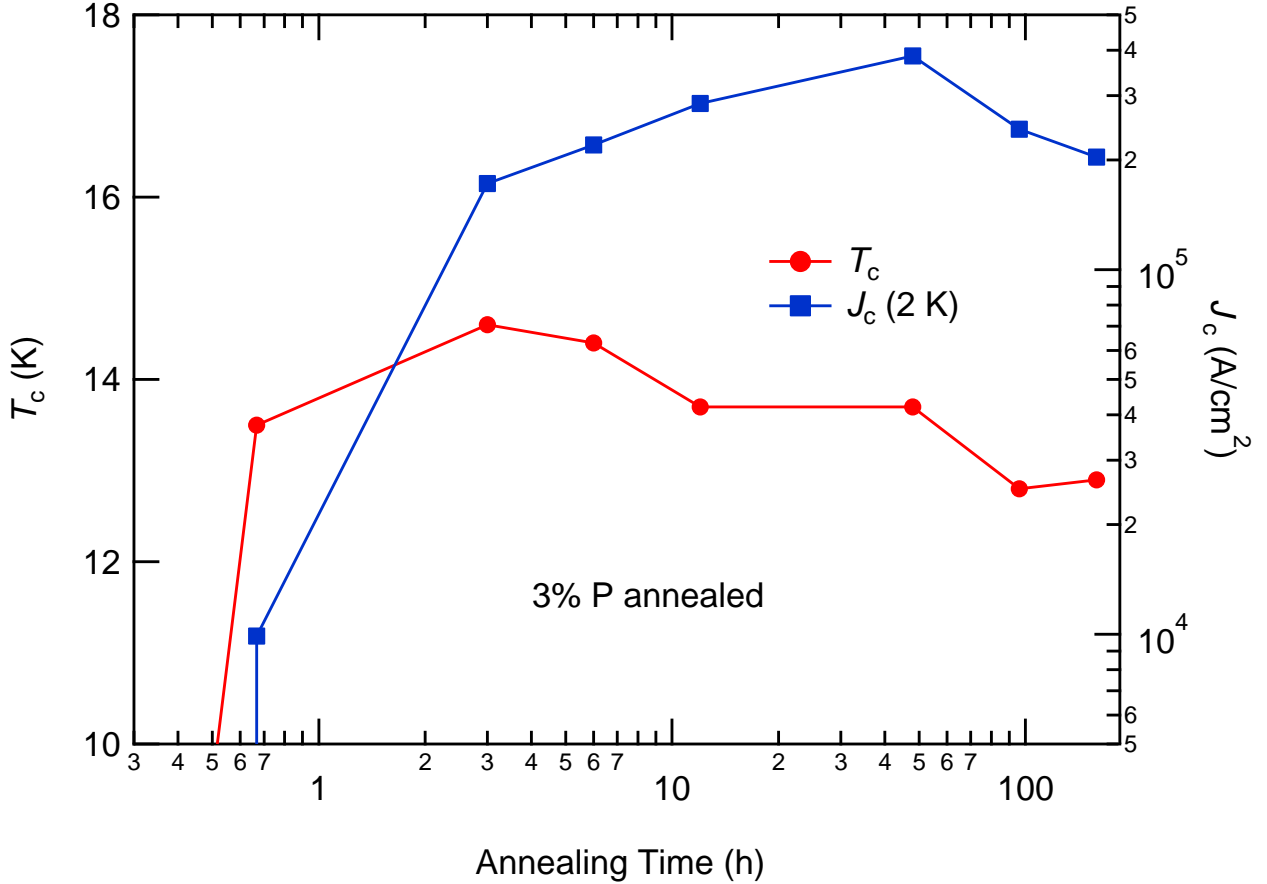}\\
\caption{(color online) Annealing time dependence of $T_\u{c}$ and self-field $J_\u{c}$ ($2~$K) for Fe$_{1+y}$Te$_{0.6}$Se$_{0.4}$ annealed in  3{\%} P atmosphere.}
\label{different_time}
\end{center}
\end{figure}
In order to obtain insight into the reason why the maximum $T_\u{c}$ for the P-annealed crystals is lower than $14~$K, as shown in Fig. \ref{TcJc}, we annealed Fe$_{1+y}$Te$_{0.6}$Se$_{0.4}$ in a $3${\%} P atmosphere for different times. 
The results are shown in Fig. \ref{different_time}.
When the annealing time is 3 h,  $T_\u{c}$ reaches the highest value of $14.6~$K.
On the other hand, $J_\u{c}$ continues to increase up to 48 h annealing.
Similar behavior of $T_\u{c}$ and $J_\u{c}$ as a function of annealing time has been observed in the case of oxygen annealing\cite{SciRep.4.4585}.
One important observation here is that $T_\u{c}$ of over $14.5~$K can be obtained even in the case of P annealing when the annealing time is appropriate. 
Thus we can claim that $T_\u{c}$ can be over $14~$K regardless of the pnictogen used for annealing.
When the annealing time is longer than 48 h, both $T_\u{c}$ and $J_\u{c}$ decrease because not only the excess iron but also the iron in Fe(Te/Se) planes reacts.

As stated above, the optimal molar ratio is different for different pnictogens.
To study the reason why the optimal molar ratio is different, we investigated the reaction products on the surface layers of the annealed crystals by XRD. 
\begin{figure}[htbp]
\begin{center}
\includegraphics[width=120mm]{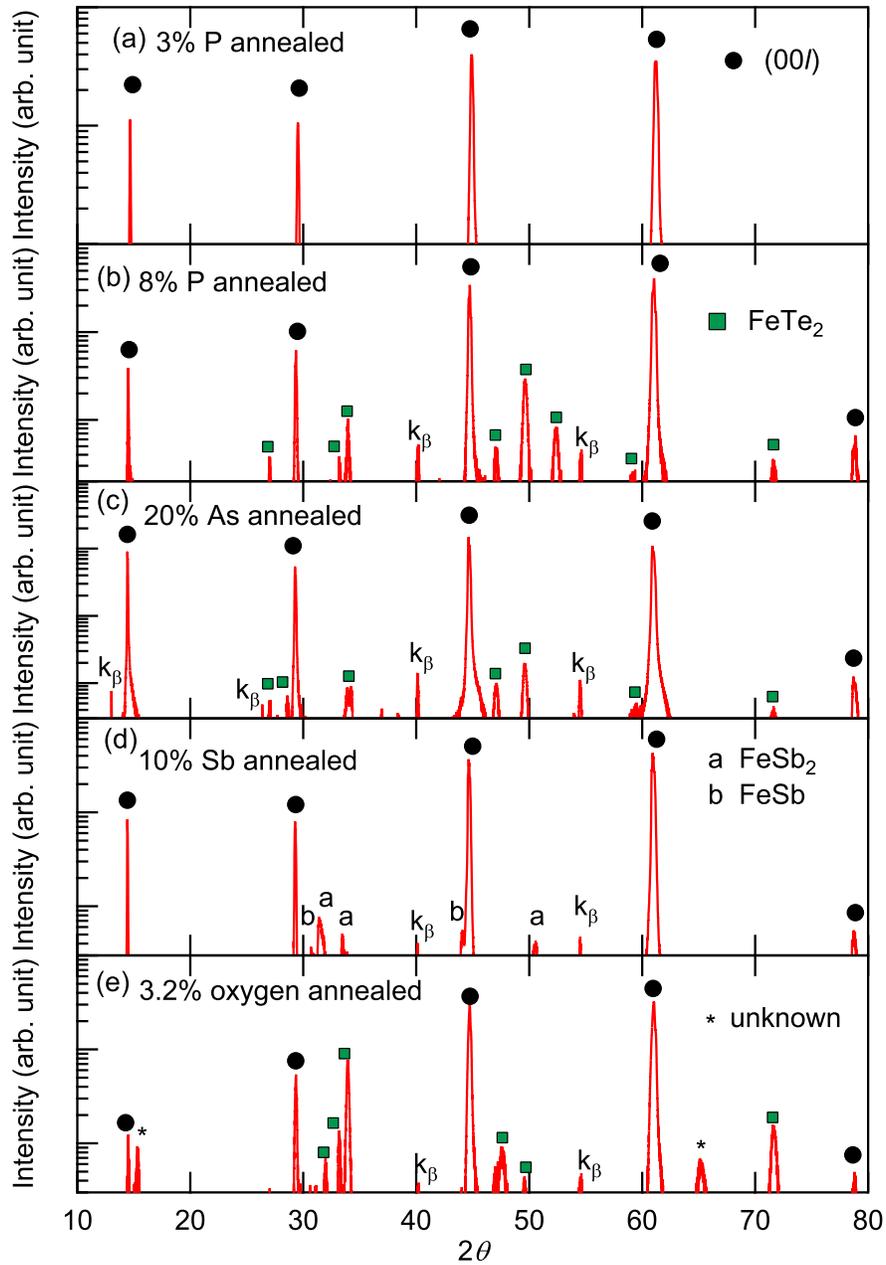}\\
\caption{(color online) XRD patterns of Fe$_{1+y}$Te$_{0.6}$Se$_{0.4}$ single crystals annealed at 400$~^\circ$C in (a) P, (b) As, (c) Sb, and  (d) oxygen atmospheres for 48 hours.}
\label{PnictogenXRD}
\end{center}
\end{figure}
Figure \ref{PnictogenXRD}(a) shows the XRD pattern of Fe$_{1+y}$Te$_{0.6}$Se$_{0.4}$ annealed at 400$~^\circ$C in a 3{\%} P atmosphere.
Only peaks of the Fe$_{1+y}$Te$_{0.6}$Se$_{0.4}$ crystal can be detected.
On the other hand, several impurity peaks can be identified in the crystal annealed in a 8{\%} P atmosphere [Fig. \ref{PnictogenXRD}(b)] because more P can easily react with the crystal.
As in the case of oxygen annealing, a large amount of FeTe$_2$ is detected in the case of 8{\%} P annealing. 
FeTe$_2$ is also detected in the As-annealed crystal. 
This result is surprising because there are no reaction products including pnictogen elements such as P and As.
However, in the case of As annealing, reaction products containing As peeled off from the surface of the crystal as mentioned above.
Thus, the original reaction products cannot be detected easily.
On the other hand, we did not detect any reaction products inside the quartz tube after P annealing.
The crystals annealed in 3{\%} and 8{\%} P atmospheres show bulk superconductivity, as can be seen in Fig. \ref{Tccompare}, indicating that excess irons existing in the crystal must be removed and present in some form.
The following scenario is possible.
Excess iron may be removed from the crystal by forming FeTe$_2$ or other materials that cannot be easily detected by XRD rather than compounds consisting of Fe and P. 
The same situation could occur in the case of oxygen annealing because very strong peaks of FeTe$_2$ comparable to the intensity of the $(00l)$ peak from the Fe$_{1+y}$Te$_{0.6}$Se$_{0.4}$ single crystal are detected, suggesting that a large amount of FeTe$_2$ exists on the surface. 
In the case of 3{\%} P annealing, FeTe$_2$ cannot be detected by XRD.
This fact suggests that FeTe$_2$ may exist inside the crystal.
In contrast to the P and As annealing, it is quite natural that binary compounds of Sb and Fe are detected [Fig. \ref{PnictogenXRD}(d)] in the case of Sb annealing.
To investigate the reaction in more detail, we also performed XRD measurements of polycrystalline powders prepared by pulverizing single crystals after annealing in different P atmospheres.
\begin{figure}[htbp]
\begin{center}
\includegraphics[width=\hsize]{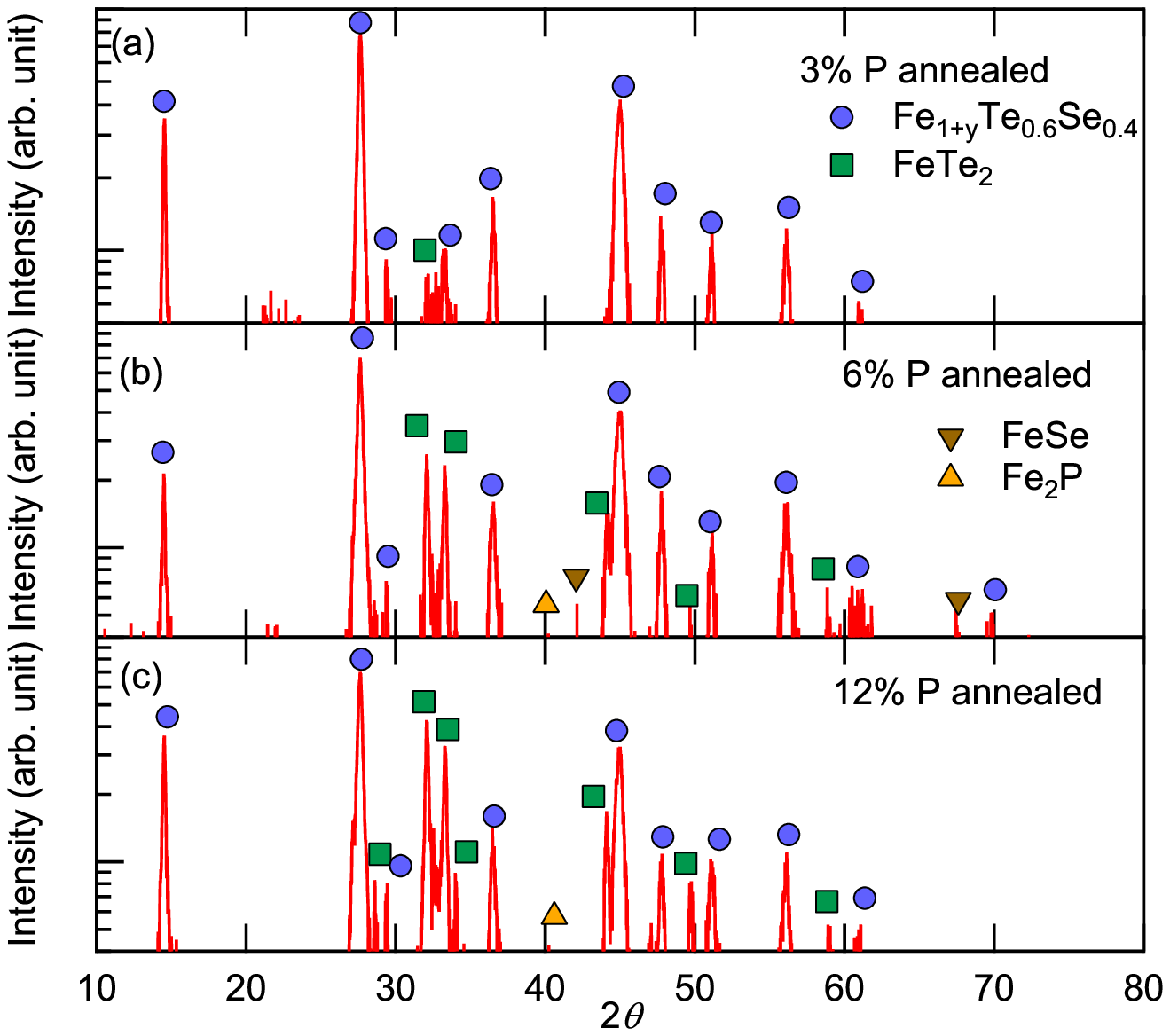}\\
\caption{(color online) XRD patterns of Fe$_{1+y}$Te$_{0.6}$Se$_{0.4}$ powders prepared by pulverizing single crystals annealed in (a) 3{\%}, (b) 6{\%}, and (c) 12{\%} P atmospheres.}
\label{PolyXRD}
\end{center}
\end{figure}
In contrast to the crystal annealed in a 3{\%} P atmosphere [Fig. \ref{PnictogenXRD}(a)], the XRD pattern of the pulverized powder shows peaks of FeTe$_2$ [Fig. \ref{PolyXRD}(a)], which demonstrates that FeTe$_2$ exists inside the crystal.
Note that bulk superconductivity appears in the case of 3{\%} P annealing, which is close to the optimal molar ratio [Fig. \ref{Tccompare}(a)].
This fact indicates that excess iron existing in the crystals is removed, which is consistent with the result that FeTe$_2$ can be detected in the case of 3{\%} P annealing.
As the molar ratio of P increases, FeSe and Fe$_2$P start to be detected in addition to FeTe$_2$, although the intensity of FeTe$_2$ is much stronger than that of other impurities.
In particular, the intensity of Fe$_2$P is very weak, suggesting that P may only serve as a catalyst rather than reacting with the excess iron directly.
This assumption is also supported by the fact that, as can be seen from Fig. 2, $T_\u{c}$ reaches $\sim 10~$K even when the molar ratio of P is only 0.5{\%}.
a similar situation also occurs in the case of As and oxygen annealing because FeTe$_2$ is detected.
In other words, As and oxygen can also serve as a catalyst in that only a small amount can remove a larger amount of excess iron.
In particular, in Ref. 24, it has been reported that oxygen can induce bulk superconductivity even when the molar ratio is only 0.5{\%}, similar to the case of P annealing.
However, Sb can only react with the excess iron directly because only binary compounds of Sb and Fe are detected.

\begin{figure}[htbp]
\begin{center}
\includegraphics[width=\hsize]{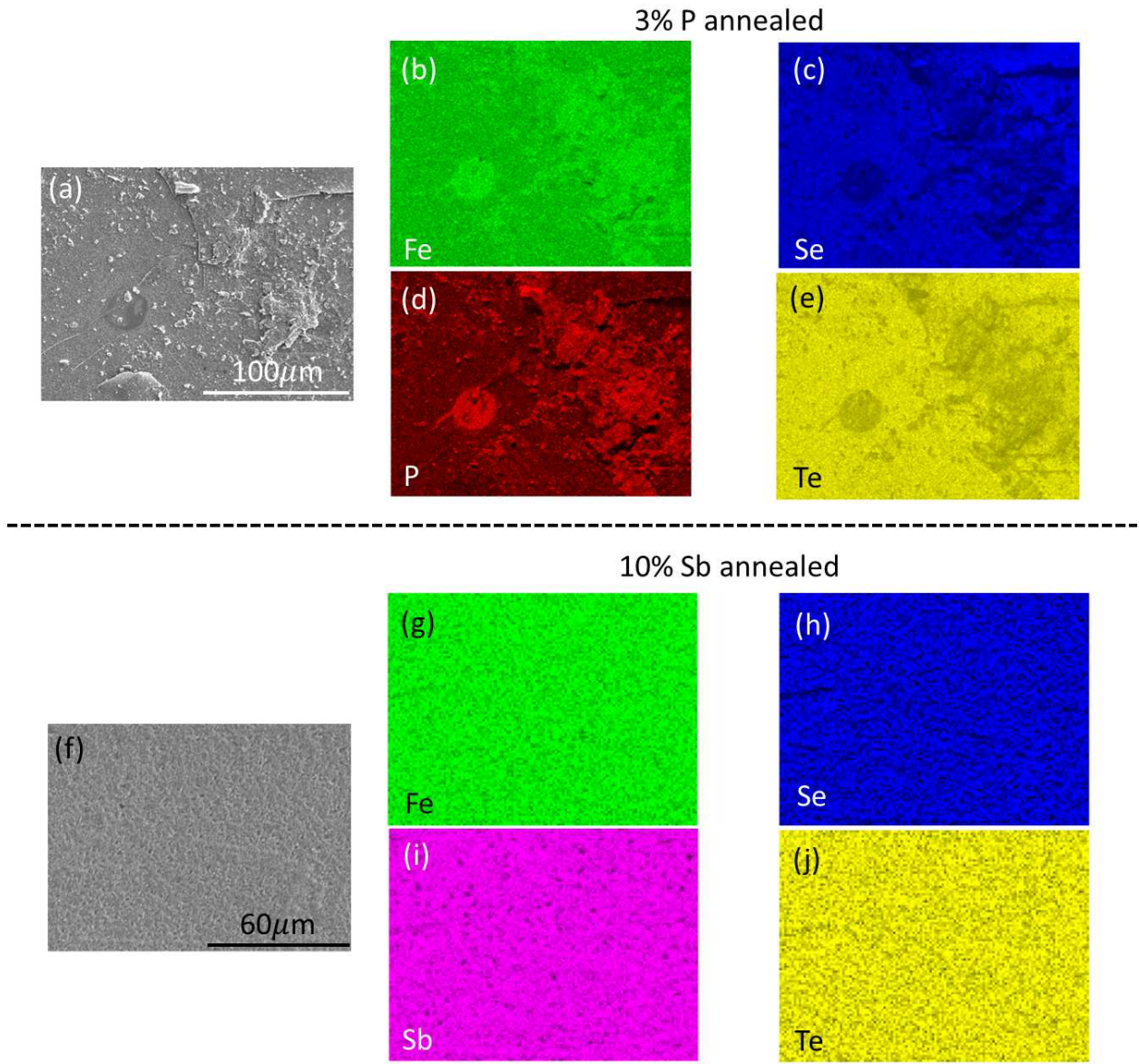}\\
\caption{(color online) (a) SEM image of Fe$_{1+y}$Te$_{0.6}$Se$_{0.4}$ annealed in 3{\%} P atmosphere for 48 h, and the distributions of (b) Fe, (c) Se, (d) P, and (e) Te probed by EDX analysis. (f) SEM image of Fe$_{1+y}$Te$_{0.6}$Se$_{0.4}$ annealed in 10{\%} Sb atmosphere for 48 h, and the distributions of (g) Fe, (h) Se, (i) Sb, and (j) Te probed by EDX analysis.}
\label{SEM}
\end{center}
\end{figure}
We also performed EDX analysis on crystals annealed in P and Sb atmospheres.
As shown in the SEM images in Fig. \ref{SEM}, the surface of the Sb-annealed crystal [Fig. \ref{SEM}(f)] is much smoother than that of the P-annealed crystal [Fig. \ref{SEM}(a)].
Figures \ref{SEM}(b)-\ref{SEM}(e) show elemental mappings of the crystal annealed in a 3{\%} P atmosphere.
Here, the distributions of Fe, Se, P, and Te are inhomogeneous and the region rich in Fe is also rich in P, indicating the formation of Fe$_2$P, although Fe$_2$P cannot be detected by XRD.
We also obtained similar results for crystals annealed in 6{\%} and 12{\%} P atmospheres.  
This inhomogeneous reaction may be related to the difficulty in obtaining a sample with $T_\u{c}$ more than $14~$K.
On the other hand, the reaction process proceeds homogeneously in the case of Sb annealing as shown in Figs. 6(g)-6(j). 
The difference in homogeneity can be ascribed to the difference in the reaction rates of the elements with Fe$_{1+y}$Te$_{0.6}$Se$_{0.4}$.

Judging from the XRD results, P can serve as a catalyst, as mentioned above.
If P serves as a catalyst and helps the excess iron to form FeTe$_2$, bulk superconductivity can be induced without forming binary compounds of P and Fe.
On the other hand, in the case of As annealing, both FeTe$_2$ and compounds consisting of Fe and As are detected.
This fact means that As has two roles in removing excess iron: one is as a catalyst to form FeTe$_2$ and the other is to react with iron directly.
In terms of the annealing mechanism, it can be considered that As is an element intermediate between P and Sb or Te, which is also supported by the fact that the optimal value is about 3-6{\%} in the case of As annealing, which is between the values in for P and Sb or Te annealing.
When the amount of P is less than or equal to the optimal value, only excess iron forms FeTe$_2$, which resides in the crystal.
When the amount of P increases to over the optimal value, the irons in Fe(Te/Se) planes in the crystal starts to form FeTe$_2$ in addition to the excess iron, resulting in the formation of a large amount of FeTe$_2$.
The fact that Te in the Fe(Te/Se) planes is consumed in the formation of FeTe$_2$ indicates that the corresponding Se should exist in the crystal.
The formation of FeSe, which is detected in the XRD patterns, is consistent with this scenario.
However, it is not clear whether all the Se is consumed to form FeSe because the intensity of the FeSe peak is very weak.

\section{Summary}
We studied the effect of annealing in pnictogen atmospheres on inducing bulk superconductivity in Fe$_{1+y}$Te$_{0.6}$Se$_{0.4}$.
Crystals annealed in P, As, and Sb atmospheres showed bulk superconductivity at a certain molar ratio of the pnictogen to the crystal. 
The highest values of $T_\u{c}$ and $J_\u{c}$ were $\sim 14~$K and 2-4$\times 10^5~$A/cm$^2$ ($2~$K, self-field), respectively, comparable to those for oxygen-annealed crystals.
However, the optimal annealing time was different for different elements due to different SVPs. 
When the crystal was annealed in P and As atmospheres, a considerable amount of FeTe$_2$ was formed as a reaction product and the optimal molar ratio was about 3-6{\%}, similar to the case of oxygen annealing. 
On the other hand, the optimal molar ratio was about 10{\%} in the case of Sb annealing, similar to the case of Te annealing.
P, As, and oxygen may serve as catalysts rather than forming binary compounds with excess iron to remove it from the crystal and induce superconductivity.

\section*{Acknowledgement}
Y. S. gratefully appreciates the support from Japan Society for the Promotion of Science.

\end{document}